# Generation of a reflected wave in an inhomogeneous medium

M.I. Ayzatsky

National Science Center Kharkiv Institute of Physics and Technology (NSC KIPT), 61108, Kharkiv, Ukraine

E-mail: mykola.aizatsky@gmail.com

In this work we present the results of calculation of the electric field distribution in the inhomogeneous media based on the generalized WKB method. In this approach the field is represented as the sum of two components, one of which is associated with the right-travelling WKB solution and the other with the left-travelling WKB solution. It has been shown that the second component is not a left-traveling nonuniform wave. Calculation results showed that the phase distribution of this component has an increasing character, similar to the phase distribution of the right-travelling wave component. Obtained results can be useful during the development of dielectric accelerating waveguides. The obtained results may also indicate the location where a semiclassical particle with energy greater than the barrier height can reflect from a smooth potential barrier.

## 1. INTRODUCTION

To describe electromagnetic fields in slowly varying inhomogeneous media the generalized WKB method is often used (see, for example, [1,2,3]). In this method the electromagnetic field is represented as the sum of two components with unknown amplitudes $\vec{E} = \vec{E}_1 + \vec{E}_2 = C_1 \vec{E}_1^{WKB} + C_2 \vec{E}_2^{WKB}$, $\vec{H} = \vec{H}_1 + \vec{H}_2 = C_1 \vec{H}_1^{WKB} + C_2 \vec{H}_2^{WKB}$ where $\vec{E}_i^{WKB}$ and $\vec{H}_i^{WKB}$ are WKB solutions. Substituting such representations in Maxwell equations yields a coupling differential system of equations for $C_1$ and $C_2$. It was assumed that component $\vec{E}_1 = C_1 \vec{E}_1^{WKB}$ ($\vec{H}_1 = C_1 \vec{H}_1^{WKB}$) is similar to a wave travelling in the positive direction (the right-travelling wave, direct, incident), and $\vec{E}_2 = C_2 \vec{E}_2^{WKB}$ ($\vec{H}_2 = C_2 \vec{H}_2^{WKB}$) is similar to a wave travelling in the negative direction ( left-traveling wave, return, reflected) [1]. This conclusion was based on the general analysis, not on the calculations.

Recently it has been proposed to use a uniform basis for description of non-periodic structured waveguides and a new generalization of the theory of coupled modes has been constructed [4,5]. Within the single-mode approximation, the fields are represented as the sum of two components, one of which is associated with the right-travelling eigenwave and the other with the left-travelling eigenwave. It has been shown that the second component is not a left-traveling nonuniform wave [6,7]. Calculation results showed that the phase distribution of this component has an increasing character, similar to the phase distribution of the right-travelling wave component. This is not an obvious result for the coupled equations. There arose a question if this behavior occurs in the case of wave coupling in inhomogeneous media.

Based on direct calculations, the properties of two components used in the generalized WKB method for describing slowly varying inhomogeneous media are studied. This paper presents the results of such a study.

## 2. MAIN EQUATIONS

We consider the simplest case of one-dimensional inhomogeneity along the z-axis and TE wave. Then the wave equation for the $E_x$ component writes as

$$\frac{d^2 E_x}{dz^2} + \frac{\omega^2}{c^2} \tilde{\varepsilon}(z) E_x = 0 , \qquad (1)$$

where $\tilde{\varepsilon}(z) = \left( \varepsilon(z) - \frac{c^2}{\omega^2} k_y^2 \right) = \left( \varepsilon(z) - \sin^2 \theta \right)$.

The solution of this equation can be represented as the sum of two new functions [8]

$$E_x = E_x^+ + E_x^- . \qquad (2)$$

By introducing two unknown functions $E_x^+$ and $E_x^-$ instead of the one $E_x$, we can impose an additional condition. This additional condition we write in the form

$$\frac{dE_x}{dz} = g^+(z) E_x^+ + g^-(z) E_x^- , \qquad (3)$$

where $g^+(z)$ and $g^-(z)$ ( $g^+(z) \neq g^-(t)$ ) are arbitrary continuous functions having continuous derivatives.

As $g_1(t) \neq g_2(t)$, from (1), (2), and (3) we can find the derivatives of $\dfrac{dE_x^+}{dz}$ and $\dfrac{dE_x^-}{dz}$

---

[1] Sometimes they are called forward and backward waves. But these terms are more often used for waves with coinciding and opposite directions of phase and group velocities.

$$\left(g^+ - g^-\right)\frac{dE_x^+}{dz} = -E_x^+\left(\frac{dg^+}{dz} + f_0 + g^+g^-\right) - E_x^-\left(\frac{dg^-}{dz} + f_0 + \left(g^-\right)^2\right),$$
$$\left(g^+ - g^-\right)\frac{dE_x^-}{dz} = E_x^+\left(\frac{dg^+}{dz} + f_0 + \left(g^+\right)^2\right) + E_x^-\left(\frac{dg^-}{dz} + f_0 + g^+g^-\right),$$
(4)

where $f_0 = \frac{\omega^2}{c^2}\tilde{\varepsilon}$.

If we choose

$$\left(g^{\pm}\right)^2 + f_0 = 0,$$ (5)

then the system (4) takes the form

$$\frac{dE_x^+}{d\xi} = \left(i\sqrt{\tilde{\varepsilon}(z)} - \frac{1}{4\tilde{\varepsilon}}\frac{d\tilde{\varepsilon}}{d\xi}\right)E_x^+ + \frac{1}{4\tilde{\varepsilon}}\frac{d\tilde{\varepsilon}}{d\xi}E_x^-,$$
$$\frac{dE_x^-}{d\xi} = -\left(i\sqrt{\varepsilon(z)} + \frac{1}{4\tilde{\varepsilon}}\frac{d\tilde{\varepsilon}}{d\xi}\right)E_x^- + \frac{1}{4\tilde{\varepsilon}}\frac{d\tilde{\varepsilon}}{d\xi}E_x^+$$
(6)

where we introduced a dimensionless variable $\xi = z\frac{\omega}{c}$.

Introducing new unknowns

$$E_x^{\pm} = \tilde{E}_x^{\pm}\exp\left[\int_{\xi_0}^{\xi}\left(\pm i\sqrt{\varepsilon(\xi')} - \frac{1}{4\varepsilon}\frac{d\varepsilon}{d\xi'}\right)d\xi'\right] = \tilde{E}_x^{\pm}\frac{1}{\sqrt[4]{\varepsilon}}\exp\left[\int_{\xi_0}^{\xi}\left(\pm i\sqrt{\varepsilon(\xi')}\right)d\xi'\right],$$ (7)

we get

$$\frac{d\tilde{E}_x^+}{d\xi} = q^+(\xi)\tilde{E}_x^-,$$
$$\frac{d\tilde{E}_x^-}{d\xi} = q^-(\xi)\tilde{E}_x^+,$$
(8)

where $q^{\pm}(\xi) = \frac{1}{4\tilde{\varepsilon}}\frac{d\tilde{\varepsilon}}{d\xi}\exp\left[\mp\int_{\xi_0}^{\xi}\left(i2\sqrt{\tilde{\varepsilon}(\xi')}\right)d\xi'\right]$.

The field representation used above is the same as in the genderized WKB method [1,2,3].

Let us consider the case when between two homogeneous half-spaces with constant permittivities $\varepsilon_l$ and $\varepsilon_r$ there is an inhomogeneous layer of length $d$. Let us also assume that in the left semi-infinite medium a wave with unit amplitude ($\tilde{E}_x^+ = 1$, $\xi < 0$) propagates towards the layer under consideration. The boundary conditions for this case have the form

$$\left.\begin{array}{l}q^+\tilde{E}_x^+ + q^-\tilde{E}_x^- = 1 + R \\ \sqrt{\tilde{\varepsilon}(0)}\left(q^+\tilde{E}_{x,1}^+ - q^-\tilde{E}_{x,1}^-\right) = \sqrt{\tilde{\varepsilon}_1}\left(1 - R\right)\end{array}\right\}, \xi = 0,$$
$$\left.\begin{array}{l}q^+\tilde{E}_x^+ + q^-\tilde{E}_x^- = T \\ \sqrt{\tilde{\varepsilon}(\rho)}\left(q^+\tilde{E}_x^+ - q^-\tilde{E}_x^-\right) = \sqrt{\tilde{\varepsilon}_2}T\end{array}\right\}, \xi = \rho,$$
(9)

where $\rho = 2\pi\frac{d}{\lambda}$, $R$ and $T$ are reflection and transmission coefficients. $\tilde{\varepsilon}_1 = \varepsilon_l - \sin^2\theta$, $\tilde{\varepsilon}_2 = \varepsilon_r - \sin^2\theta$.

A general method of numerical analyzing the wave propagation and scattering problem for inhomogeneous planar-layered media is the finite difference method (see, for example, [9,10,11]). We used more accurate method based on the 4th order Runge-Kutta method [12]. The peculiarity of the system (8) is the presence of an integral with a variable upper limit. Therefore, to implement the Runge-Kutta method with a step $h$ it is necessary to split the interval [0, $\rho$] with a step $h/4$ and use the Simpson formular to calculate the values of the integral with a step $h/2$.

The final algebraic system of equation we write as

$$q_1^+ \tilde{E}_{x,1}^+ + q_1^- \tilde{E}_{x,1}^- = 1 + R,$$

$$\sqrt{\tilde{\varepsilon}(0)} \left( q_1^+ \tilde{E}_{x,1}^+ - q_1^- \tilde{E}_{x,1}^- \right) = \sqrt{\tilde{\varepsilon}_1} \left( 1 - R \right),$$

$$\left. \begin{array}{l} \tilde{E}_{x,s}^+ = \alpha_{1,1}(s) \tilde{E}_{x,s-1}^+ + \alpha_{1,1}(s) \tilde{E}_{x,s-1}^-, \\ \tilde{E}_{x,s}^- = \alpha_{2,1}(s) \tilde{E}_{x,s-1}^+ + \alpha_{2,2}(s) \tilde{E}_{x,s-1}^-, \end{array} \right\} \quad s = 2,\ldots,N \qquad (10)$$

$$q_N^+ \tilde{E}_{x,N}^+ + q_N^- \tilde{E}_{x,N}^- = T,$$

$$\sqrt{\tilde{\varepsilon}(\rho)} \left( q_N^+ \tilde{E}_{x,N}^+ - q_N^- \tilde{E}_{x,N}^- \right) = \sqrt{\tilde{\varepsilon}_2} T,$$

where $q_s^\pm = q^\pm \{h s\}, E_{x,s}^\pm = E_x^\pm \{h s\}, \ s = 1, N$.

## 3. CALCULATION RESULTS

To study the properties of the two components $E_x^\pm$ used in the generalized WKB method, the distributions of these functions on the interval $[0, d]$ are calculated for several laws of change of permittivity in a layer: constant, linear and one in which the first and the second derivatives equal to zero at the ends of the interval $z = 0, d$

$$\varepsilon_I(z) = \begin{cases} \varepsilon_l, z < 0, \\ \varepsilon_r, z \geq d, \end{cases} \qquad (11)$$

$$\varepsilon_{II}(z) = \begin{cases} \varepsilon_l, z < 0, \\ \varepsilon_l + (\varepsilon_r - \varepsilon_l) \dfrac{z}{d}, & 0 \leq z \leq d, \\ \varepsilon_r, z > d, \end{cases} \qquad (12)$$

$$\varepsilon_{III}(z) = \begin{cases} \varepsilon_l, z < 0, \\ \varepsilon_l + (\varepsilon_r - \varepsilon_l) \left\{ 10 \dfrac{z^3}{d^3} - 15 \dfrac{z^4}{d^4} + 6 \dfrac{z^5}{d^5} \right\}, & 0 \leq z \leq d, \\ \varepsilon_r, z > d. \end{cases} \qquad (13)$$

We considered the case when $\varepsilon_l = 1$, $\varepsilon_r = 10$, $d/\lambda = 3$, $\theta = 0$.

Dependencies of $\varepsilon$ on $z/\lambda$ are presented in Figure 1.

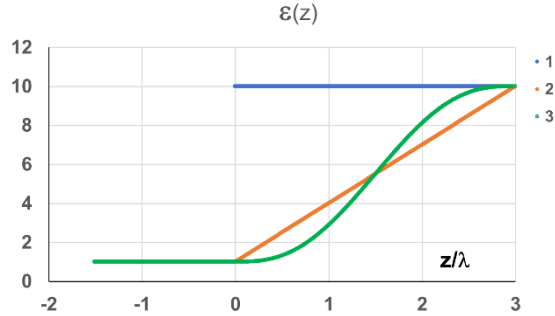

*Figure 1 Dependency of $\varepsilon$ on $z/\lambda$: 1 - $\varepsilon = \varepsilon_I$, 2- $\varepsilon = \varepsilon_{II}$, 3- $\varepsilon = \varepsilon_{III}$.*

Results of calculations are given in Figure 2-Figure 7 ($h = d/500$). For the case of reflection from a homogeneous dielectric (see Figure 2, Figure 3) we obtain the expected results. The reflection coefficient $R = 0.519493853295916$ does not differ from the strict one $\left( \sqrt{\varepsilon_r} - \sqrt{\varepsilon_l} \right) / \left( \sqrt{\varepsilon_r} + \sqrt{\varepsilon_l} \right) = 0.519493853295916$. The reflected wave is generated in the plane $z = 0$ and, therefore, exists at $z < 0$.

For the case of reflection from an inhomogeneous dielectric with the layer which permittivity changes linearly with the coordinate (Figure 4, Figure 5) the component $E_x^+$ is similar to a wave travelling in the positive direction ($d\varphi^+ / dz > 0$) in the entire space $-\infty < z < \infty$. The component $E_x^-$ exists at $-\infty < z < d$. It is similar to a wave travelling in the negative direction ($d\varphi^- / dz < 0$) at $-\infty < z < 0$, but at $0 < z < d$ it is similar to a nonuniform wave travelling in the positive direction ($d\varphi^- / dz > 0$). The reflection coefficient for this case is much smaller $R = 5.45E-2$. Reducing the step size by a factor of three changes the reflection coefficient by $0.02E-2$.

To calculate $\tilde{E}_x^+, \tilde{E}_x^-$ in the first approximation [1,2,3] it is necessary to set $q^+(\xi) = 0$ ($\tilde{E}_x^+ = const$) The results of such calculations show that the field distribution remains virtually unchanged.

For more smother transition (Figure 6, Figure 7) the situation is the same. The reflection coefficient for this case is 3.71E-3. The component $E_x^+$ is similar to a wave travelling in the positive direction and its amplitude is a smooth function of coordinate. The behavior of component $E_x^-$ do not change, exclude the location of the plane where the derivative of phase of $E_x^-$ change its sign. It is shifted into the nonuniform layer.

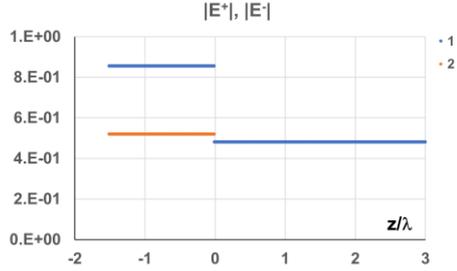

*Figure 2 Dependences on $z/\lambda$ of the amplitude of the components of electric fields: 1- $|E_x^+|$, 2 - $|E_x^-|$ at $\varepsilon = \varepsilon_I$*

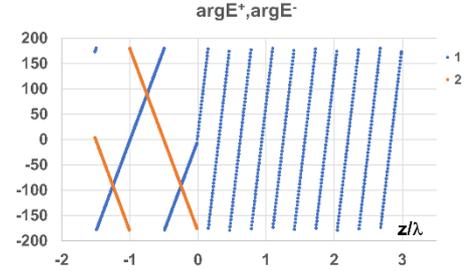

*Figure 3 Dependences on $z/\lambda$ of the phase of the components of electric fields: 1- $\arg E_x^+$, 2 - $\arg E_x^-$ at $\varepsilon = \varepsilon_I$*

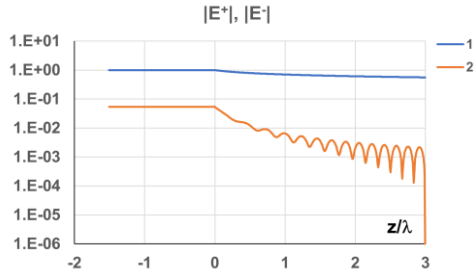

*Figure 4 Dependences on $z/\lambda$ of the amplitude of the components of electric fields: 1- $|E_x^+|$, 2 - $|E_x^-|$ at $\varepsilon = \varepsilon_{II}$*

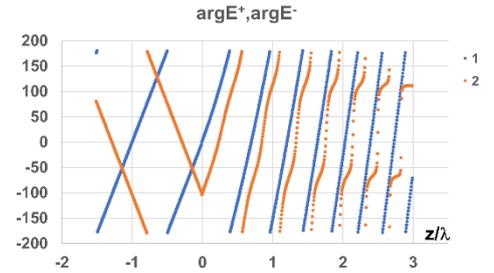

*Figure 5 Dependences on $z/\lambda$ of the phase of the components of electric fields: 1- $\arg E_x^+$, 2 - $\arg E_x^-$ at $\varepsilon = \varepsilon_{II}$*

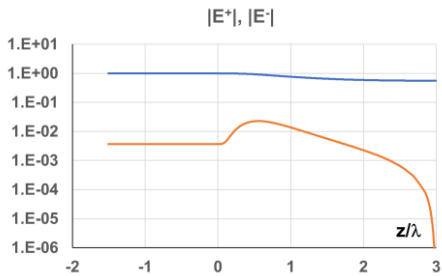

*Figure 6 Dependences on $z/\lambda$ of the amplitude of the components of electric fields: 1- $|E_x^+|$, 2 - $|E_x^-|$ at $\varepsilon = \varepsilon_{III}$*

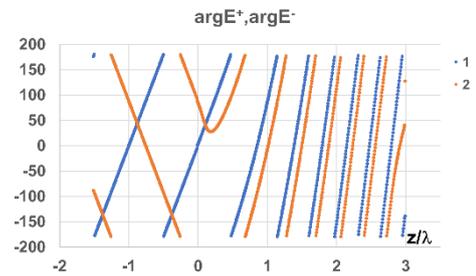

*Figure 7 Dependences on $z/\lambda$ of the phase of the components of electric fields: 1- $\arg E_x^+$, 2 - $\arg E_x^-$ at $\varepsilon = \varepsilon_{III}$*

Electromagnetic power flow per unit area perpendicular to the $z$ axis in our designations is

$$P = \frac{1}{2}\operatorname{Re}(E_x H_y^*) = \frac{1}{2Z_0}\sqrt{\tilde{\varepsilon}(z)}\left(|E_x^+|^2 - |E_x^-|^2\right) = \frac{1}{2Z_0}\left(|\tilde{E}_x^+|^2 - |\tilde{E}_x^-|^2\right) = \frac{1}{2Z_0}Y, \qquad (14)$$

where $Z_0 = \sqrt{\mu_0/\varepsilon_0}$, $|\tilde{E}_x^+(0)|^2 = 1$.

Since we are considering lossless media, the parameter $Y$ must be constant along the $z$ axis. Calculations show that the difference $Y(0)-Y(\rho)$ is small. It is of the order of $2\times 10^{-4}$ for $h=6\pi/500$ and $6\times 10^{-5}$ for $h=6\pi/1500$.

## 4. CONCLUSIONS

In this work we present the results of calculation of the electric field distribution in the inhomogeneous media based on the generalized WKB method. In this approach the field is represented as the sum of two components, one of which is associated with the right-travelling WKB solution and the other with the left-travelling WKB solution. It has been shown that the second component is not a left-traveling nonuniform wave. Calculation results showed that the phase distribution of this component has an increasing character, similar to the phase distribution of the right-travelling wave component.

Dielectric-loaded accelerating structures offer the potential of a simple, inexpensive alternative to copper disk-loaded structures for use in high-gradient rf linear accelerators (see, for example, [13,14,15]). According to the tendency of developing the copper structures, the nonuniform dielectric structure will be more efficient than the homogeneous ones. As in accelerating structures the field distribution plays the crucial role, obtained results can be useful during the development of dielectric accelerating waveguides.

As the equation (1) is the simplest form of the Schrödinger equation, the obtained results may also indicate the location where a semiclassical particle with energy greater than the barrier height can reflect from a smooth potential barrier.